# Collapse of superconductivity in a hybrid tin-graphene Josephson junction array

Zheng Han[1,2], Adrien Allain[1,2], Hadi Arjmandi-Tash[1,2], Konstantin Tikhonov[3,4], Mikhail Feigel'man[3,5], Benjamin Sacépé[1,2] and Vincent Bouchiat[1,2]

[1] *Univ. Grenoble Alpes, Institut NEEL, F-38042 Grenoble, France.*
[2] *CNRS, Institut NEEL, F-38042 Grenoble, France.*
[3] *L. D. Landau Institute for Theoretical Physics, Kosygin street 2, Moscow 119334, Russia*
[4] *Dept. of Condensed Matter Physics, The Weizmann Institute of Science, 76100 Rehovot, Israel.*
[5] *Moscow Institute of Physics and Technology, Moscow 141700, Russia.*

**When a Josephson junction array is built with hybrid superconductor/metal/superconductor junctions, a quantum phase transition from a superconducting to a two-dimensional (2D) metallic ground state is predicted to happen upon increasing the junction normal state resistance. Owing to its surface-exposed 2D electron gas and its gate-tunable charge carrier density, graphene coupled to superconductors is the ideal platform to study the above-mentioned transition between ground states. Here we show that decorating graphene with a sparse and regular array of superconducting nanodisks enables to continuously gate-tune the quantum superconductor-to-metal transition of the Josephson junction array into a zero-temperature metallic state. The suppression of proximity-induced superconductivity is a direct consequence of the emergence of quantum fluctuations of the superconducting phase of the disks. Under perpendicular magnetic field, the competition between quantum fluctuations and disorder is responsible for the resilience at the lowest temperatures of a superconducting glassy state that persists above the upper critical field. Our results provide the entire phase diagram of the disorder and magnetic field-tuned transition and unveil the fundamental impact of quantum phase fluctuations in 2D superconducting systems.**

The ground state terminating 2D superconductivity in the superconductor-(metal)-insulator quantum phase transition remains an unsolved fundamental problem[1,2]. In granular[3,4] and some amorphous superconducting thin films[5-10], oxide interfaces[11,12] and Josephson junction arrays[13,14], an intervening metallic state is often observed experimentally at the disorder-tuned or magnetic field-tuned quantum critical point separating the superconducting from the (weak) insulating state. In contrast to usual non-interacting 2D electronic systems in which electron localization yields insulating ground state[15], this metallic state is characterized by the presence of electron-pairing fluctuations that are indicated in experiments by a partial drop of resistance — a reminiscence of the nearby superconducting phase — preceding the nearly $T$-independent resistive state.

The possibility that a 2D system can undergo a zero-temperature superconducting to metal transition is rather intriguing[16-22]. The nature of this metallic state, sometimes termed Bose-metal[18,21], where charge carriers are supposedly Cooper-pairs, is not clear. Considerable theoretical efforts have been focused on the role of quantum phase-fluctuations of the superconducting order-parameter in the regime where the superconducting phase stiffness breaks down. In particular, models involving specific types of Josephson junction arrays in which superconducting disks are coupled together via a 2D disordered metal through proximity effect have been considered[16,17,20,22]. Such proximity-coupled arrays encompass all the physical ingredients: conventional superconducting electron pairing, embedded dissipation channels in the normal disordered metal, and quantum fluctuations of superconducting phases due to weak Coulomb blockade on superconducting disks[23].

The theoretical analysis of these models shows that a quantum phase transition from a superconducting state to a non-superconducting state can be tuned by two different means: either geometrical by increasing the distance in between disks, or by increasing the metal resistance. Both parameters directly affect the superconducting proximity effect and result in a dramatic enhancement of quantum phase fluctuations. Though proximity effect is expected to extend over infinite distance at zero temperature leading systematically to a superconducting ground state, the main conclusion of recent theories[16,17,20,22] is that quantum phase fluctuations lead to a collapse of superconductivity, that is, of the array phase stiffness. This should arise at a given critical value of the tuning parameter, leading to a zero-temperature 2D metallic state.

On the experimental side, it is hardly conceivable to access this quantum breakdown of superconductivity by tuning continuously the resistance of a metallic thin film that carries the superconducting proximity effect. Varying the inter-disk distance on a large set of samples has thus remained the only reliable approach with thin films. A recent study of proximity arrays made of gold thin-films covered with an array of niobium nano-disks[24] has demonstrated that the critical temperature of such proximity coupled arrays decreases upon increasing the inter-disk distance and extrapolates to zero for a finite inter-disk distance, suggesting a quantum phase transition to a non-superconducting, metallic ground state.

In this work, we take a different route to tune the quantum phase transition in a controllable fashion. We use graphene[25], a purely 2D crystal of carbon atoms, to serve as the diffusive metal, which we decorate with a triangular array of tin disks (a type-I superconductor). The gate-tunable sheet resistance of graphene, which can approach the quantum of resistance ($h/e^2$), enables us to continuously vary the strength of disorder which affects the inter-disk proximity effect and therefore tune the transition from the superconducting to the metallic state. Furthermore, the charge-carrier density of graphene is always several orders of magnitude below those of classical metals, preventing the weakening of the superconductivity on the disks by inverse proximity effect[26]. As we describe below, these capabilities, along with quantitative comparisons with theory, allow us to unveil both the collapse of superconductivity by quantum phase fluctuations, and the ensuing 2D metallic state. In addition, under perpendicular magnetic field a resilient superconducting glass state is observed at our lowest temperatures, as expected from theory[27,28,29].

**Berezinski-Kosterlitz-Thouless transition in superconducting array**

Our proximity-induced array consists of a triangular lattice of disk-shaped superconducting islands that are deposited onto a graphene Hall bar equipped with a back-gate electrode (see Fig.1). The superconducting disks are made of 50 nanometers thick tin films, and have a diameter $2a$ = 400 nm, with a thickness of 50 nm. Disks are separated by a distance $b$ = 1 $\mu$m in between their centres (see Fig.1). The electron mobility of the graphene extracted from a fit of the field effect and from Hall measurements is about 680 cm$^2$V$^{-1}$s$^{-1}$ at high gate voltage (Fig. S1), while the mean free path $l$ can be varied with the backgate within the range of 10 to 30 nm (Fig. S2), corresponding to a diffusion coefficient $D$ = 50 to 140 cm$^2$s$^{-1}$ (see Fig. S2). Our system differs considerably from previous experiments performed on Sn-graphene hybrids[4,30] for which tin clusters of random shapes self-assembled densely on graphene surface by dewetting. Those experiments were performed in the opposite regime of short junctions (coherence length >>b) and for one of it[4] in the limit of strongly disordered graphene (mean free path << b) for which a superconductor-to-insulator quantum phase transition has been observed upon decreasing the graphene charge carrier density.

We begin the presentation of the data by describing the two-step transition[31,32] that the sample undergoes towards the superconducting state.

Upon cooling from the normal state, a gate-independent resistance drop occurs at a temperature of ~ 3.6 K (Fig. 2a, red and blue curves, also Fig. S3). We attribute this first drop to the superconducting transition of the Sn nanodisks. The resistance drop $\Delta R$ at this transition amounts to 20 % on the electron side. This number is in good agreement with the surface filling factor of ~15 % (see Fig. S3) which demonstrates the good transparency of nanodisks/graphene Andreev contacts.

Further cooling enhances the proximity effect via the graphene, which eventually leads to the percolation of superconductivity and the establishment of a 2D superconducting state (see Figs S3a & S3b). As shown in Fig.2a, at gate voltages between 0 and 30 V, the resistance decreases when temperature is lowered from 3 K to 0.06K. An opposite trend is seen for gate voltages close to Dirac point (namely, -30 V to 0 V).

Below 1 K (Fig. S3b) and for gate voltage above -3V, the proximity-coupled array develops full superconductivity with a zero resistance state and a gate-dependent critical temperature. The transition to the superconducting state in our 2D sample is captured by a Berezinski-Kosterlitz-Thouless (BKT) mechanism[33,34] that describes the superconducting transition as a proliferation and unbinding, at $T_c^{BKT}$, of vortex-antivortex pairs, that is, thermal phase fluctuations of the superconducting order parameter. A colour-scaled map showing resistance versus gate-voltage and temperature is shown in Fig. 2b. $T_c$ is indicated by the white contour. One can see that higher positive gate-voltage leads to higher $T_c$, whereas $T_c$ vanishes in the region near to the Dirac point, between -30 V and -3 V.

For Josephson junction arrays, the value of $T_c^{BKT}$ is directly proportional to the Josephson energy $E_J(b,T)$ of a single junction formed between two superconducting nano-disks. For a triangular lattice, the relation reads[29]:

$$T_c^{BKT} \simeq 1.47 E_J(b, T_c^{BKT}), \qquad (1)$$

where $E_J(b,T)$ depends only on the metal conductance, the inter-disk distance b and the diffusion coefficient D (see methods). Using experimentally extracted values of $D$ and of the graphene resistance measured at 4 K for all back-gate values, we calculated $T_c^{BKT}$ by solving (1) self-consistently with respect to $T$ for the entire back-gate range (see methods). In Fig. 2b, the resulting $T_c^{BKT}$ is shown by the dashed line. For high gate voltage, that is, $V_g > 10V$, this theoretical value of $T_c^{BKT}$ is in excellent agreement with the sample critical temperature without any fitting parameter, thereby validating our theoretical description.

The critical supercurrent at $T=0$ provides another computable physical quantity on such arrays. As shown in Fig. 3a, the plot of the differential resistance $dV/dI$ versus bias current and $V_g$ allows to obtain the full $V_g$ dependence on $I_c$ at $T= 0.06$ K. In the zero temperature limit, the critical current of an array of nanodisks can be inferred from the Josephson coupling energy of neighbouring disks[29], which is given by:

$$E_J(b, T=0) = \frac{\pi^3}{4} \frac{g\hbar D}{b^2 \ln^2(b/a)} = \frac{\hbar}{2e} I_1, \qquad (2)$$

where $g = \hbar/(e^2 R_\square)$ is the dimension-less conductance of graphene in the normal state,

with $R_\square$ the sheet resistance; and $I_1$ is the maximum supercurrent between two neighbouring nanodisks. The total critical current $I_c$ can be estimated by summing the critical current $I_1$ of each nanodisk neighbours (see Fig. 1b) of our array, which are contributing as parallel channels, and by neglecting small contribution to $I_c$ from non-nearest pairs of disks, that is, $I_c \approx 6I_1$. According to the above relations, at $T = T_c$, the product of $I_c R_N$ ($R_N$ being the graphene sheet resistance in the normal state) should only depend on diffusion coefficient $D$ for a given geometry. This can be directly tested since the use of graphene as a 2D disordered metal enables to gate-tune the diffusion coefficient while keeping all geometrical aspects of the array constant. By extracting the experimental critical current $I_c$ from Fig. 3a, and measuring $R_N$ at 4 K (Fig. 2a), we plot the quantity $eI_c R_N / \hbar D$ as a function of the gate voltage in Fig. 3b. One can see that above a given doping level this quantity reaches a constant level of $3.6\times10^9$ cm$^{-2}$ which matches the quantity $3\pi^3 / b^2 \ln^2(b/a)$ (dashed line). Therefore, at high gate voltages, this experiment follows the theoretical predictions for such proximity-coupled arrays of disk sparsely decorating graphene[29].

**Quantum breakdown of superconductivity**

The central result of this work is the anomalous reduction of the sample's critical temperature $T_c$ with respect to the calculated $T_{BKT}$ upon increasing graphene resistance, leading to a sudden collapse of superconductivity when approaching the charge neutrality point of the graphene layer (dark red region of Fig. 2b). Indeed, the $V_g$-dependence of $T_c$ shown in Fig. 2b first deviates from $T_c^{BKT}$ towards lower temperatures for $V_g$ < 10V, and then abruptly drops to zero at around -3V. The suppression of superconductivity is also apparent in the $V_g$-dependence of the critical supercurrent $I_c$ at 0.06 K. As can be seen in Fig. 3, the zero-resistance state (blue region, marked as "S" in Fig. 3a) disappears when $V_g$ = -3V in a similar fashion as for $T_c$. We attribute these significant deviations, culminating in the collapse of superconductivity, to the failure of the BKT model[29], which only take into account thermal phase fluctuations. Therefore our results are a direct consequence of the breakdown of the superconducting phase stiffness due to the emergence of quantum phase fluctuations[16,17,29].

The quantum origin of phase fluctuations in such proximity-coupled arrays stems from the particle number-phase uncertainty relation that states that when the number of particles in an isolated superconducting disk is fixed, the phase of the superconducting order parameter will undergo strong non-thermal fluctuations. In tunnelling Josephson junction arrays, such quantum phase fluctuations are expected to happen when the disk charging energy overcomes the Josephson coupling energy[13], a situation that leads to charge quantization and the Coulomb blockade of the tunnelling into the disks. Proximity-coupled arrays behave differently since their superconducting disks are well-coupled to a metal layer without tunnel barrier. When the normal-state resistance of the array approaches a few k$\Omega$, which is achievable with graphene, the Andreev conductance, that is, the conductance for electron pairs, can be such that a weak charge quantization[23] of charge $2e$ prohibits the charge transfer into the disks. As the transfer of quasiparticles is also hindered at $T = T_c$ by the superconducting gap, the total charge of the disks tends to become fixed and thus promotes quantum fluctuations of the phase of the superconducting order parameter (see Supp. Info). As a result, theory predicts that the presence of quantum phase fluctuations leads to an effective decoupling of superconducting phases between disks, and eventually to the collapse of the array phase stiffness for a given critical resistance of the 2D metal[16,17]. The wide range of tunability of graphene sheet resistance is the key ingredient that allows to access to this regime of weak charge quantization and the ensuing superconducting-metal transition.

We next investigate the nature of the state terminating superconductivity for $V_g < -3$V. We systematically measured the $T$-dependence of the resistance, that we term $R$-$T$ curve, for gate-voltage close to the critical $V_g = -3$V. Shown in Fig. 4a, the $R$-$T$ curves in the non-superconducting regime at $V_g < -3$V exhibit a resistance drop below 0.5 K reminiscent of the superconducting transition of those at $V_g > -3$V, which indicates that superconducting fluctuations have developed into the graphene. However, instead of falling into a zero-resistance superconducting state, the $R$-$T$ curves level off into finite resistance plateaus of gate-dependent values that can reach up to 28 k$\Omega$. Importantly we can exclude the possibility of electron heating as the origin of this metallic tail at low temperature, since in a magnetic field such that the tin nanodisks turn metallic the resistance shows a clear T-dependence till the base temperature (see Fig. S4).

To get more insights on this critical regime we show in Fig. 3b a set of differential resistance versus bias current curves extracted from the data of Fig. 2a for few values of $V_g$. In the superconducting state, zero resistance is observed up to the critical current as discussed before (see Fig. 3b, curve at $V_g = 0$ V). However, for $V_g < -3$ V where the R-T curves saturates at finite values upon cooling, the differential resistance around zero current bias transform into a zero-bias dips which persist even close to the charge neutrality point at $V_g = -13$ V. These dips give a clear indication that superconducting fluctuations exist for all gate voltages. Therefore we are led to conclude that the ground state of our array, which exhibits finite resistance plateaus, is superconductivity-related. Moreover, the fact that the *R-T* curves conspicuously extrapolate to a constant resistance at zero-*T* suggests that a 2D quantum metallic state is approached. We believe that the nature of this collapse of superconductivity and the resulting quantum metallic state arises from the enhancement of quantum phase fluctuations upon increasing graphene resistance, which eventually leads to the effective decoupling of the nanodisks superconducting phases. For critical temperatures below the typical Thouless energy of a single junction, $E_{Th} = \hbar D / b^2$, the impact of quantum fluctuations on $T_c$ is expected[17] to be dramatic with an exponential suppression in the form $T_c \sim E_{Th} \exp\left(-\dfrac{c}{g - g_c}\right)$, where $g_c$ is the critical conductance of the graphene and *c* a constant, which qualitatively explains the observed collapse of superconductivity. The variation of the resistance plateau in the metallic state (Vg < -3V) with gate voltage may be partially attributed to superconducting fluctuations of Aslamazov-Larkin[35] and Maki-Thompson[36,37] types which would include the phase fluctuations, leading to a positive correction to the conductance. A non-perturbative theory of these fluctuations is required to quantitatively explain the very strong contribution of superconducting fluctuations to conductivity near critical Vg = -3 V.

**Superconducting glassy state**

Another route to induce a quantum phase transition to the normal state in our superconducting array consists in applying a perpendicular magnetic field, $H$. The destruction of the array superconductivity by the magnetic field — we consider here a magnetic field smaller than the critical field of tin nanodisks — results from the interplay between three different effects. First the magnetic field imposes phase frustration between the nanodisks. This frustration, which is given by $f = HS_0/\Phi_0$ ($S_0 = \sqrt{3}b^2/2$ is the area of the array elementary cell in the triangular lattice and $\Phi_0 = h/2e$ the flux quantum), impedes to reach a set of phase differences between nano-disks which simultaneously minimises the Josephson energy of all junctions. Second, the magnitude of each individual Josephson coupling $E_J$ between two neighbouring disks $i$ and $j$ decreases upon increasing magnetic field in a way such that $E_J$ drops exponentially beyond the magnetic length $L_H = \sqrt{\Phi_0/H}$ due to random magnetic field-induced phase shifts, whereas the typical modulus of $E_J$ is suppressed by a factor $\sim 1/g$ only[38]. Thereby both these effects tend to suppress superconductivity in the array. The second effect however leads to the enhancement of quantum phase fluctuations by the mitigation of the proximity effect[17], whereas frustration may yield non-monotonic behaviour of the resistance upon increasing $H$ as observed in tunnelling Josephson junction arrays[13,14]. The third effect to consider here comes from the phase coherent nature of the disordered graphene electron gas and the resulting mesoscopic fluctuations. As a result of the multiple interferences of electron wave functions, the graphene conductance undergoes mesoscopic fluctuations that lead, in the normal state, to universal conductance fluctuations. In our proximity-coupled array, such mesoscopic fluctuations are present together with superconducting fluctuations, which result in resistance fluctuations upon varying electron density, as apparent in the gate evolution of the resistance measured at 0.06 K shown in Fig. 2a. In a similar fashion, small variations of magnetic field can randomly modulate the Andreev conductance of each Josephson junction by dephasing the electron interference pattern[27,38,39] and therefore play an important role in the setting of global phase coherence, especially when the mean critical current is strongly suppressed by magnetic field.

The conjunction of these three effects has led to the theoretical predictions that, at very low temperature, superconductivity may persist above the upper critical field, $H_{c2}$, in the form of an upturn[28,29] of the T-dependence of $H_{c2}$, or a possible re-entrant superconducting phase[27]. The resulting state is expected to be a superconducting phase glass due to the phase frustration between nanodisks and to the mesoscopic fluctuations that arbitrarily affect each Josephson coupling.

To address these fundamental aspects we systematically measured the T-evolution of the magneto-resistance of the sample in the superconducting phase far from the charge neutrality point ($V_g$ = 30 V). Fig. 5 presents the whole set of data in the form of a plot showing the colour-coded magneto-resistance as a function of $T$ and $H$. The dark blue area indicates vanishingly small resistance, that is, superconducting state, and the other colours indicate the resistive normal state. Interestingly we observe that the superconducting phase extends over a wide and continuous range of magnetic field, which resembles the mixed phase of bulk superconductors. The T-evolution of $H_{c2}$ from 0.8 K down to 0.2 K indeed follows closely the theoretical behaviour of type-II superconductor[26] that we have traced in black line in Fig. 5. At this gate voltage, the zero-T extrapolation of this theoretical fit gives *2.2* mT which is close to full frustration *f*=1 reached at the field value $H_{c0} = \Phi_0 / S_0 = 2.3 \, mT$, that is, one flux quantum per unit cell.

However, inspecting Fig. 5 for $T$ below 0.2 K, we observe that the experimental $H_{c2}(T)$ curve deviates upward and draws a resilient superconducting pocket persisting up to 3.3 mT. Replicas of different sizes, though of finite resistivity, are even visible at higher magnetic field. Importantly, in this re-entrant superconducting phase, the supercurrent is strongly suppressed in comparison with the main superconducting region at the same temperature (see Fig. S5).These observations comply with the theoretically expected superconducting phase glass state[27,28,29]. Our measurements therefore constitute a clear demonstration of reentrant superconductivity in a inhomogeneous 2D superconducting system and provide another fundamental aspect of such proximity coupled arrays.

**Phase diagram of the quantum superconductor-to-metal transitions**

The use of graphene has provided here a unique opportunity to explore in a single device the phase diagram in phase-space variables $T$, $H$ and $V_g$, where $V_g$ is equivalent to disorder, which drive the thermal and the quantum phase transitions to the normal state. Due to the continuous tuning of disorder through the gate-tuneable graphene electron density, we can therefore construct the entire phase diagram of our proximity-coupled JJ array. The result shown in Fig. 6 provides the complete picture of the experimental temperature, disorder and magnetic field-driven superconductor-to-metal transitions addressed in this work.

The similarities with the original phase diagram based on the dirty-boson model[40] are striking and show that our proximity coupled array behaves in many respects as disordered superconducting thin films. However, the new physical ingredient here relies on the inhomogeneous nature of superconductivity, which presents built-in superconducting disks subject to phase fluctuations. Such inhomogeneities also emerge close to the quantum critical point in amorphous thin films[41] in the form of self-induced fluctuations of the superconducting state induced by disorder [42].

To conclude, our work demonstrates that by increasing the normal metal resistance in a proximity-coupled array, a metallic state can terminate the superconducting state due to a strong enhancement of quantum phase fluctuations, which dramatically suppress the critical temperature $T_c$. The use of graphene as 2D electron gas platform has enabled us to continuously map the entire disorder and magnetic field-driven superconductor-to-metal transition. This plot has let us to observe the appearance of a reentrant superconducting phase glass state under magnetic field above the predicted critical field. This system offers new insights to unveil the dramatic impact of quantum fluctuations induced by weak charge quantization in 2D superconducting systems.

# Methods

We used CVD-grown graphene sheets transferred onto 285 nm oxidized silicon wafer as 2D diffusive metal. As shown in Fig. 1, the sample was patterned by standard e-beam lithography into Hall bar geometry (central square area 6 μm$^2$) and contacted with normal leads (50nm Au/5nm Ti). The entire graphene surface was then decorated with a second lithography step by an array of 50-nm-thick Sn nanodisks.

Samples were anchored to the mixing chamber of a $^3$He-$^4$He dilution refrigerator placed inside a shielded cryostat and connected to highly filtered lines. Lossy coaxial lines, capacitive filters, and π-filters were installed at the mixing chamber stage to attenuate the electronic noise from about 10 kHz. A low-frequency lock-in measurement setup based on current biased method (2 to 5nA input current) was used to measure the differential resistance while voltage across the sample was measured with low-noise preamplifiers.

To fit the $T_{BKT}$ in the measured data in Fig. 2b, we replaced out Eq.(1) with the $E_J$ in the left side of the Eq. (5) in Ref. 29, which is the Josephson energy derived from the collective Josephson coupling in a 2D array using Matsubara-space Usadel equation, with the result:

$$E_J(b,T) = 4\pi g T \sum_{\omega_n > 0} \left\{ \pi / [2\ln(\sqrt{D/2\omega_n}/a)] \right\}^2 P(\sqrt{\omega_n/2E_{Th}}) . \qquad (a)$$

The above equation, together with (1), can be reduced to

$$6\pi g \sum_0^\infty \frac{\pi^2}{\ln^2(\hbar D/2a^2\omega_n)} P(\sqrt{\omega_n/2E_{Th}}) = 1, \qquad (b)$$

where $E_{Th} = \frac{\hbar D}{b^2}$ is the Thouless energy of the pair of disks separated by distance $b$, $a$ the diameter of superconducting disks, $\omega_n = \pi T(2n+1)$ is the $n^{th}$ Matsubara frequency at temperature $T$. The function $P(z) = z \int_0^\infty K_0(z\cosh t) K_1(z\cosh t) dt$, with $K_n(x)$ the MacDonald-Bessel function. Eq. (b) contains only one variable, which is the critical temperature $T$ at the BKT transition, since the parameters of $g$ and $D$ can be obtained from measured data. By summing up (b) (high orders of $n$ do not affect the sum significantly since the function $P$ decays rapidly with increasing $n$), one gets the dashed line as the unique solution of Eq. (b) from experimentally extracted $g$ and $D$.

**Acknowledgements**. Samples were fabricated at the NANOFAB facility of the Néel Institute, the technical team of which has been of critical help for this work. We gratefully thank D. Shahar for valuable discussions and comments on the manuscript. We thank N. Bendiab, H. Bouchiat, C. Chapelier, J. Coraux, C.O. Girit, B. M. Kessler, L. Marty, A. Reserbat-Plantey, and A. Zettl for stimulating discussions. This work is financially supported by ANR-BLANC projects SuperGraph, TRICO and Cleangraph. DEFI Nano ERC Advanced Grant MolNanoSpin. Z.H. and H.A-T acknowledge grant support from Cible program of Région Rhone-Alpes and from Nanosciences foundation respectively. Research of M.F. was partially supported by the RFBR grant #13-02-00963.

# Figure captions

**Figure 1. Proximity-coupled array of superconducting nanodisks on graphene.** (a) Schematics of the device involving a triangular array of superconducting tin disks decorating the bare graphene surface. The device connected to a current source and measured using a 4-probe measurement setup. (b) Optical micrograph of the sample. The black lines underline the graphene sample plasma etched hall bar, which is connected to gold electrodes seen on the edge of the image. The graphene surface is decorated with a regular array of tin nanodisks of diameter 400nm separated by 1µm between centres. (c) Scanning electron micrograph showing the triangular lattice of tin disks. (Scale bars in both images are 1 micrometer).

**Figure 2. Collapse of superconductivity in the proximity coupled array.**
(a) Gate voltage dependence of the four-probe resistance for three different temperatures. The curve at T = 4K gives the gate-dependence of the device resistance above the tin nanodisks superconducting transition. At T=3K, the resistance drops over the full gate range due to the superconducting transition of the tin nanodisks. At the base temperature of 0.06 K, the array undergoes a transition from superconducting to resistive state for -29 V<$V_g$<-3 V and the resistance exhibits a sharp peak at the charge neutrality point. (b) Color-scaled map of resistance versus temperature and gate voltage. The superconducting phase (S) on the electron-doped side is shown by the blue area and the superconducting critical temperature bordering the resistive state is indicated by the white contour. The dashed line is calculated from Eq. (1) without fitting parameter. The experimental $T_c$ deviates around $V_g$ = 5 V towards a sharp breakdown of the superconducting state at $V_g$ = -3 V.

**Figure 3. Critical current in the proximity coupled array at 0.06 K.** (a) Color-scaled map of the differential resistance dV/dI measured at 0.06 K versus gate voltage and bias current. The border between the zero-resistance state (blue area) and the resistive state (red area) indicates the critical current $I_c$. (b) The quantity $eI_c R_N / \hbar D$ calculated from data of (a) and from Fig. S2 and plotted as a function of gate voltage. The red dotted line corresponds to the theoretical value calculated with eq. (2).

**Figure 4. Zero-temperature metallic state at criticality.** (a) Temperature dependence of array resistance measured at different gate voltages in a temperature range of 0.07 K – 0.5 K. While the resistance drop into a full superconducting state for $V_g$ > -5V, the superconducting transition is incomplete for -29 V<$V_g$ < -5V and a T-independent resistance plateau extends till the base temperature of 0.06 K. (b) Individual current-biased differential resistance measured at 0.06 K for a set of gate voltages ranging from -13 V to 0 V.

**Figure 5. Reentrant superconductivity under magnetic field.** Color-scaled map of magnetoresistance plotted as a function of temperature and magnetic field measured in the superconducting phase ($V_g$ = 30 V). The upper critical field of the array is indicated by the white contour that borders the superconducting and resistive state. The solid line is a fit of the upper critical field by the BCS theory for dirty superconductors. A clear deviation happens below 0.2 K where a re-entrant pocket of superconductivity persists up to 3.3 mT and is followed at higher field by not fully superconducting pockets, non periodic in magnetic field.

**Figure 6. Phase diagram of the superconductor-to-metal transition**
A 3D-phase diagram showing the superconducting state reconstructed from three color-scaled maps of the array resistance in ($V_g,H,T$) space. The resistance in ($V_g$-$H$) space is measured at 0.06 K. As the gate voltage $V_g$ has a direct action on the effective disorder of the graphene mediating superconductivity, this axis has been relabelled accordingly. Notice the superconducting reentrance above the first critical field which can be seen in both *($V_g,H$)* and *(H,T)* planes.

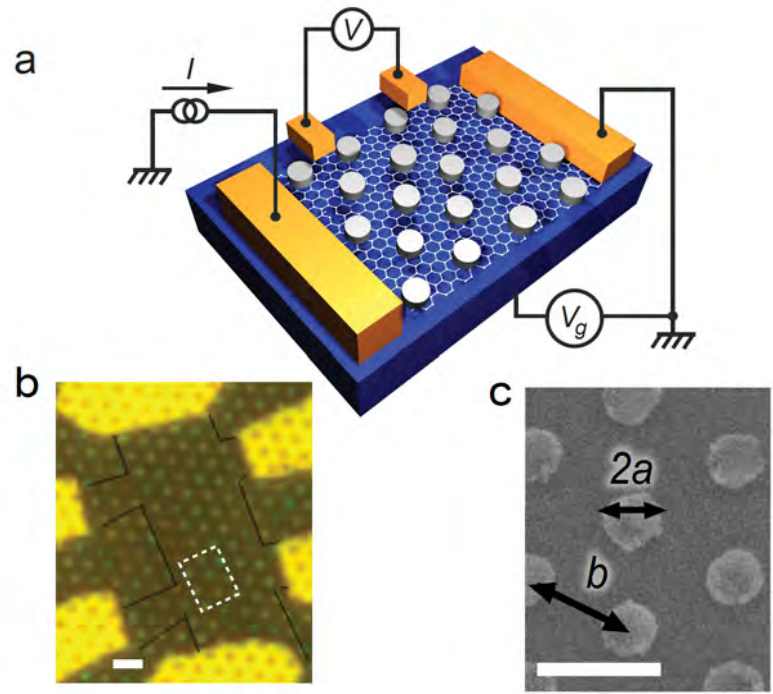

**Figure 1 Han *et al.***

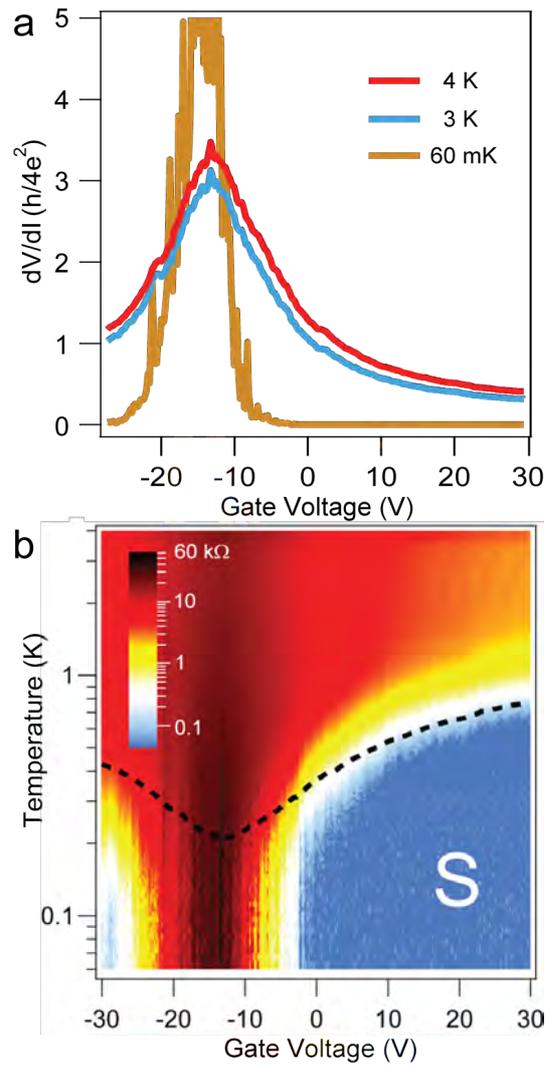

**Figure 2 Han *et al.***

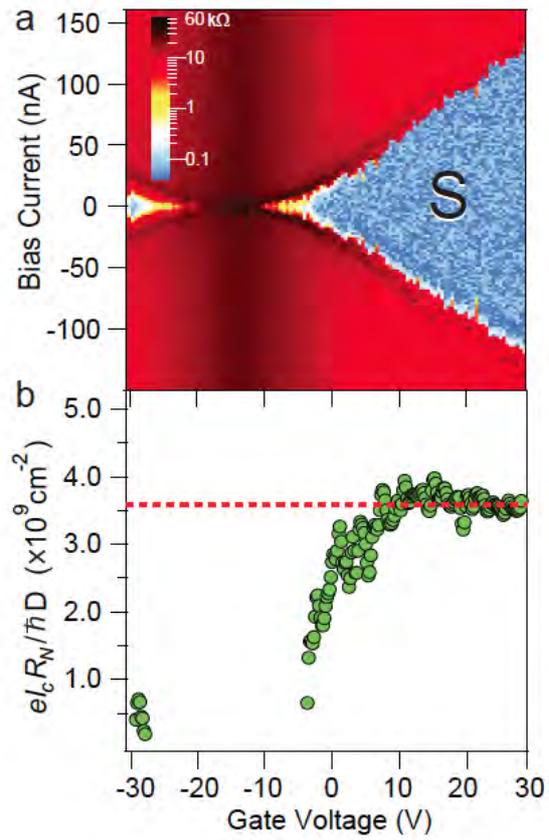

**Figure 3 Han *et al.***

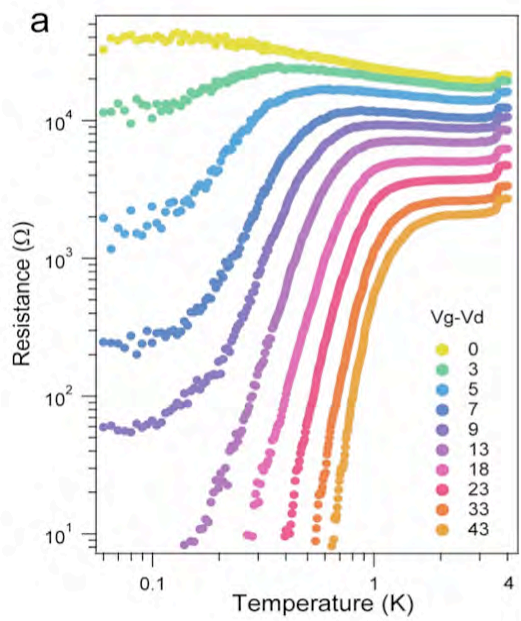

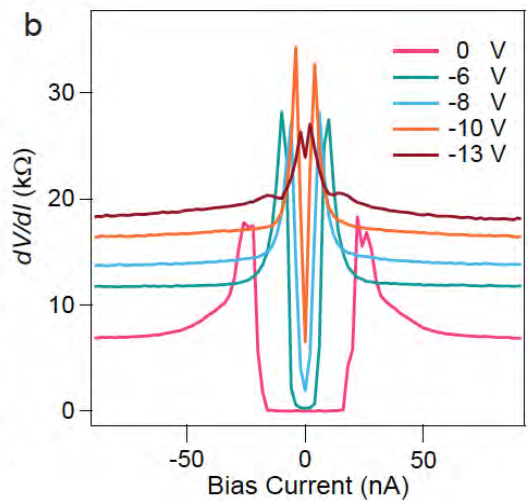

**Figure 4 Han *et al.***

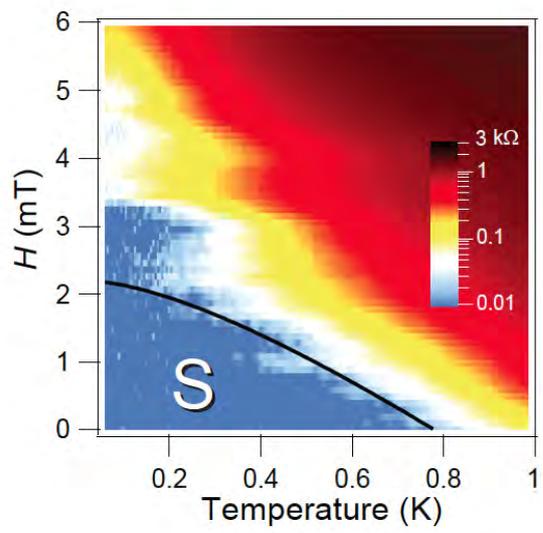

**Figure 5 Han *et al.***

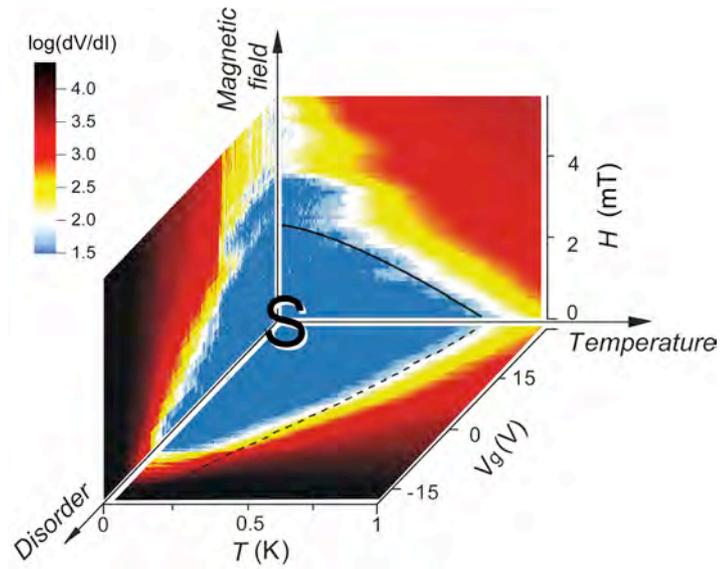

**Figure 6 Han *et al.***